\documentclass{aa501}
\usepackage{graphicx}
\begin{document}
%
%\thesaurus{03         % A&A Section 3: Astronomical Instrumentation ...
%          (03.20.7;   % Techniques : radial velocities
%	   03.09.6;   % Instrumentation : spectrographs
%	   08.15.1;   % Stars : oscillations
%          08.16.2)}  % Stars : planetary systems

\title{P-mode observations on $\alpha$~Cen~A\thanks{Based on observations
collected with the CORALIE echelle spectrograph on the 1.2-m Euler Swiss
telescope at La Silla Observatory, ESO Chile}}
\author{F. Bouchy and F. Carrier} 

\offprints{F. Bouchy}
\institute{Observatoire de Gen\`eve, 51 ch. des Maillettes, 1290 Sauverny, 
           Switzerland \\}

\date{Received 9 May 2001 / Accepted 5 June 2001}

\authorrunning{Bouchy \& Carrier}
\titlerunning{P-mode observations on $\alpha$~Cen~A}

\abstract{
We have made a clear detection of p-mode oscillations in the nearest solar-like 
G2V star $\alpha$~Cen~A with the CORALIE spectrograph on the 1.2-m Swiss telescope at 
the ESO La Silla Observatory. We report the 5~nights of observation on this 
star in May 2001 during which 1260 high precision radial velocity measurements 
were obtained. The power spectrum clearly shows several identifiable 
peaks between 1.7 and 3~mHz. The average large splitting of 105.7~$\mu$Hz and 
the amplitude of about 35 cm\,s$^{-1}$ of these modes are in agreement with 
theoretical expectations. 
\keywords{Stars: individual: $\alpha$~Cen~A --
	  Stars: oscillations --
	  Techniques: radial velocities}
}

\maketitle

\def\cms{\,cm\,s$^{-1}$}      %cm s-1
\def\ms{\,m\,s$^{-1}$}        %ms -1
\def\kms{\,km\,s$^{-1}$}      %kms -1
\def\vsini{$v$\,sin\,$i$}     %vsini
\def\m2s2{\,m$^{2}$\,s$^{-2}$} %m2 s-2

\section{Introduction}

One of the major objectives of the stellar physic is the understanding of 
stellar interiors and the modelisation of star evolution. Asteroseismology, 
which consists to measure amplitudes and frequencies of oscillation modes is 
an ideal tool to constrain stellar models and evolutionary 
theory.

Recent results of ground-based observations using Doppler techniques have 
permitted to detect solar-like oscillations in the F5IV-V star Procyon (Martic 
et al. \cite{martic99}) and the G2IV star $\beta$ Hydri (Bedding et al. 
\cite{bedding01}; Carrier et al. \cite{carrier01}). These two stars exhibit 
good evidence for excess power centered at 1 mHz with peak amplitude of 
about 50 \cms.

A primary and challenging target for the search for p-mode oscillations is the 
G2V star Rigil Centaurus ($\alpha$~Cen~A, HR\,5459). Thanks to its proximity 
and its belonging to a binary system, its characteristics are well-determined which
simplifies the interpretation of asteroseimological results. Scaling from the 
solar case, the frequency of its greatest mode is expected to be 
$\nu_{\rm max}\,=\,2.3$ mHz, the primary frequency splitting
$\Delta\nu_{\rm 0}\,=\,105.8$~$\mu$Hz and the oscillation amplitude 
$A_{\rm osc}\,=\,31.1$ {\cms} (Kjeldsen \& Bedding \cite{kjeldsen95a}). Detailed 
Models with seismological analysis have been recently proposed for 
$\alpha$~Cen~A (see, e.g., Pourbaix et al. \cite{pourbaix99}, Guenther \& 
Demarque \cite{guenther00}, Morel et al. \cite{morel00}).

Several groups have already conducted thorough attempts to detect signature of
the p-mode oscillations on this star. Two groups claimed mode detections
with amplitudes 3.2\,-\,6.4 greater than solar (Gelly et al. \cite{gelly86}; 
Pottasch et al. \cite{pottasch92}). It was however infirmed by three others 
groups which obtained upper limits of mode amplitudes of 1.4\,-\,3 times solar 
(Brown \& Gilliland \cite{brown90}; Edmonds \& Cram \cite{edmonds95}; 
Kjeldsen et al. \cite{kjeldsen99}). More recently Schou \& Buzasi (\cite{schou01}) 
have made photometric observations of $\alpha$~Cen~A with the WIRE spacecraft 
and reported a possible detection of P-modes.

We obtained high precision radial velocity (RV) measurements on $\alpha$~Cen~A 
and report in this paper our observations and the power spectrum analysis of 
this star.

\section{Observations and data reduction}

The observations of $\alpha$~Cen~A were carried out with the CORALIE fiber-fed 
echelle spectrograph (Queloz et al. \cite{queloz00}) mounted on the 1.2-m Swiss telescope 
at the ESO La Silla Observatory. CORALIE is the southern hemisphere twin of 
the ELODIE spectrograph (Baranne et al. \cite{baranne96}), both of them well known for 
their discoveries of extra-solar planets. The wavelength domain 
ranges from 3875 to 6820 {\AA} recorded on 68 orders. Thanks to a slightly 
different optical combination at the entrance of the spectrograph and the use 
of a 2k by 2k CCD with 15-$\mu$m pixels, CORALIE reaches a spectral 
resolution of 50\,000 with a 3 pixels sampling. The total efficiency including
atmosphere, telescope, spectrograph and detector was measured and reaches about 
1.5~\% at 5500~{\AA} in the best seeing case. During stellar exposures, the 
spectrum of a thorium lamp 
carried by a second fiber is simultaneously recorded in order to monitor the 
spectrograph instability and thus to obtain high precision velocity 
measurements. Spectra are extracted and reduced in real time using the
INTER-TACOS software package developed by D. Queloz and L. Weber at the Geneva
Observatory (see Baranne et al. \cite{baranne96}). 

The standard RV computation by a cross-correlation algorithm was replaced by an
algorithm based on the optimum weight procedure first proposed by Connes 
(\cite{connes85}).
We developed this method and showed that the optimum procedure is a factor 1.6 
more efficient in term of photon noise than the cross-correlation procedure 
(Bouchy et al. \cite{bouchy01}). The method uses the full spectral information
available and compares two spectra of the same source to compute their velocity
difference. Its present limitation is that the Doppler shift 
must remain small compared to the line-width. This is the case in the time 
scale of few hours when the velocity change, dominated by Earth motion, is 
less than 100~{\ms} ($\sim$~1/50 of a typical stellar line-width). During a whole 
night, Earth motion can introduce a Doppler shift greater than 500 \ms. An 
error of about 3 to 5 \% is then introduced if an unique ''reference'' spectrum 
is used during the whole night. However, this effect can be entirely eliminated by 
a low order polynomial fit subtraction without consequences for the 
characterization of high frequency oscillation modes.

\begin{figure}
\resizebox{\hsize}{!}{\includegraphics{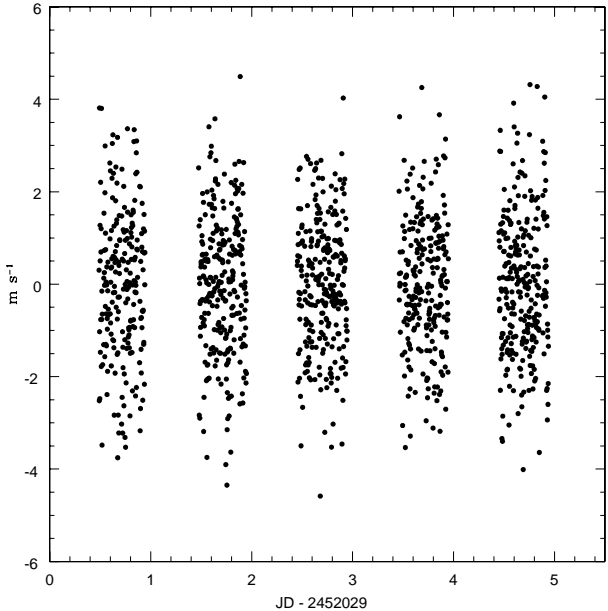}}
\caption{Radial velocity measurements of $\alpha$~Cen~A. A 3 order 
polynomial fit subtraction was applied on each night. The dispersion 
reaches 1.47 \ms.}
\end{figure}

$\alpha$~Cen~A was observed over 5 nights in May 2001. A journal of these
observations is given in Table 1. We took sequences of 40-s exposures with a
dead time of 110-s in-between. In total, 1260 spectra were collected 
with typical signal-to-noise ratio in the range 300\,-\,420 at 550 nm. 

\begin{table}
\caption{Distribution and dispersion of RV measurements}
\begin{center}
\begin{tabular}{cll}
\hline
Date & Nb spectra & $\sigma$ (\ms) \\ \hline
2001/04/29 & 220 & 1.58 \\
2001/04/30 & 259 & 1.49 \\
2001/05/01 & 268 & 1.39 \\
2001/05/02 & 246 & 1.40 \\
2001/05/03 & 267 & 1.53 \\ \hline
\end{tabular}
\end{center}
\end{table}

Radial velocities are computed for each night relatively to the highest 
signal-to-noise ratio spectrum obtained in the middle of the night. The 
resulting velocities are presented in Fig. 1. The dispersion of 
these measurements reaches 1.47 {\ms} and the individual value of each night 
is listed in Table 1. The precision of our RV measurements is compared with 
the fundamental uncertainty due to photon noise using the method described by 
Bouchy et al. (\cite{bouchy01}). The uncertainties coming from the stellar 
spectrum and from the simultaneous thorium spectrum used in the instrumental 
tracking are typically the same and equal to 0.50 \ms. The quadratic sum of these 
two photon noise contributions is 0.70 \ms.

\section{Power spectrum analysis}

In order to compute the power spectrum of the velocity time series of Fig. 1, 
we used the Lomb-Scargle modified algorithm (Lomb \cite{lomb76}, Scargle 
\cite{scargle82}) for unevenly spaced data. The resulting LS periodogram, 
shown in Fig. 2, exhibits a series of peaks between 1.7 and 3 mHz modulated 
by a broad envelope, which is the typical signature of solar-like oscillations. 
This signature also appears in the power spectrum of every individual night. 
Toward the lowest frequencies ($\nu$ $<$ 0.5 mHz), the power rises and scales 
inversely with frequency squared as expected for instrumental instabilities.  
The mean white noise level $\sigma_{ps}$, computed in the range 0.5-1.5 mHz, 
reaches $3.06\times 10^{-3}$ \m2s2. With 1260 measurements, the velocity accuracy 
corresponds thus to 0.98 \ms. 

\begin{figure}
\resizebox{\hsize}{!}{\includegraphics{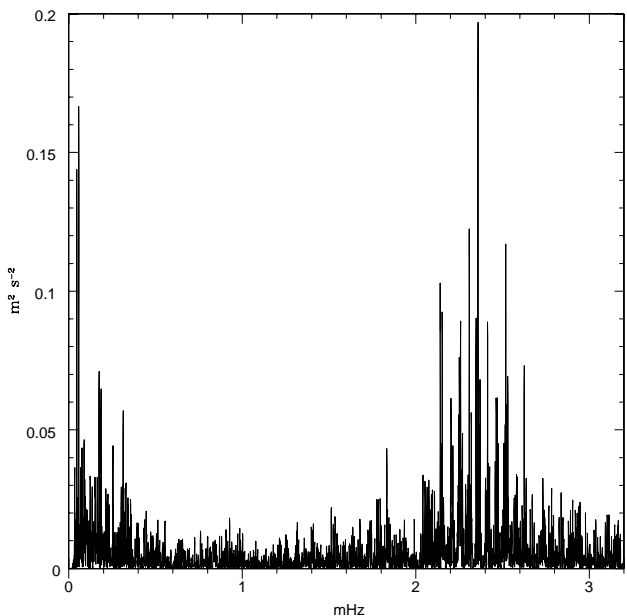}}
\caption{Power spectrum of the radial velocity measurements of 
$\alpha$~Cen~A}
\end{figure}

Observed mode frequencies expected for solar-like stars correspond to low angular degree 
$l$ and high radial order $n$ and are well approximated by the asymptotic 
relation :\\

\begin{eqnarray}
\label{eq1}
\nu(n,l) & \approx & \Delta\nu(n+\frac{l}{2}+\epsilon)
\end{eqnarray}   
with $\Delta\nu\,=\,\nu_{n,l}-\nu_{n-1,l}$\,.\\
\\
The large splitting $\Delta\nu$ reflects the average stellar density and 
$\epsilon$ is a constant near unity (1.46 for the Sun) sensitive to the 
surface layers. 
In order to determine the average large splitting 
$\langle \Delta\nu\rangle$, we calculate the comb response of the power spectrum as 
proposed by Kjeldsen et al. (\cite{kjeldsen95b}). An average comb response 
is computed using all peaks greater than 0.03~{\m2s2} (i.e. with 
S/N greater than 3 in the amplitude spectrum) in the 
frequency range 1.7\,-\,3~mHz. The resulting comb response shown in Fig. 3 gives 
$\langle \Delta\nu\rangle\,=\,105.7$~$\mu$Hz. 

\begin{figure}
\resizebox{\hsize}{!}{\includegraphics{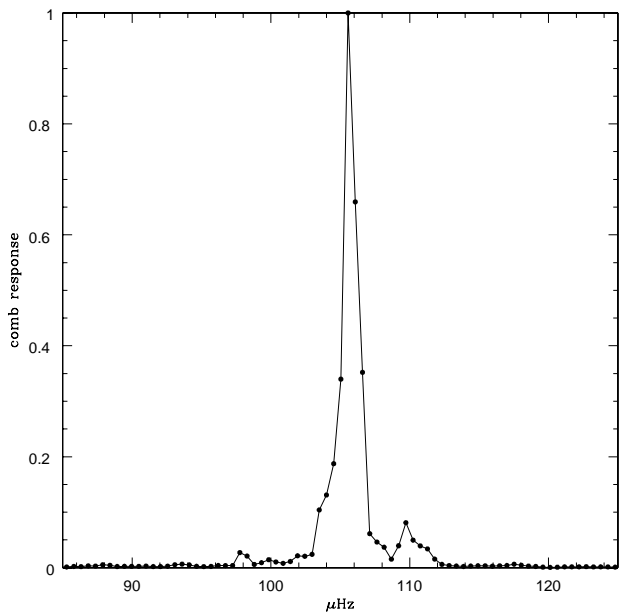}}
\caption{Average comb response of the power spectrum of $\alpha$~Cen~A 
computed with 0.5 $\mu$Hz resolution}
\end{figure}

In order to estimate the average amplitude of modes, we followed the method
described and used by Kjeldsen \& Bedding (\cite{kjeldsen95a}). A simulated 
time series was generated with artificial signal plus noise using our window
function. The generated signal corresponds to 20 modes centered at 2.36 mHz, 
separated by $\Delta\nu/2$ and convolved by a gaussian envelope. The generated 
noise is gaussian with a standard deviation of 1 \ms. Several simulations show 
that the average amplitude of peaks is in the range 33\,-\,37 \cms.

An attempted identification of the strongest modes was processed with the 
CLEAN algorithm (Roberts et al. \cite{roberts87}) in order to remove the 
effect of the window function. The identified modes are listed in Table~2 and 
shown in Fig.~4. The values of $n$ and $l$ are deduced from the 
asymptotic relation (see Eq.~\ref{eq1}) assuming that the main peak
corresponds to $n\,=\,21$ and $l\,=\,0$. With this configuration, the parameter 
$\epsilon$ is estimated to 1.35. This value is located in-between the Sun
value (1.46) and the value deduced from theoretical models (Guenther \& 
Demarque \cite{guenther00}, Morel et al. \cite{morel00}) which give $\epsilon$ 
in the range 1.07\,-\,1.27. The last row in Table~2 gives the average large splitting for $l\,=\,0$ and 
$l\,=\,1$. Both values are in agreement with this deduced from the comb 
response.

\begin{table}
\caption{Mode frequencies (in $\mu$Hz) of $\alpha$~Cen~A estimated with 
the CLEAN procedure. The frequency resolution of our time series is 2.6 $\mu$Hz.}
\begin{center}
\begin{tabular}{ccc}
\hline 
 & $l$ = 0 & $l$ = 1 \\ \hline
n=16 & 1833.1 & - \\
n=17 & - & - \\
n=18 & 2041.1 & 2095.6 \\
n=19 & 2146.4 & 2203.1\\
n=20 & 2251.0 & 2308.4 \\
n=21 & 2358.9 & 2414.1 \\
n=22 & 2463.3 & 2519.1 \\
n=23 & 2566.5 & 2625.1 \\
n=24 & 2671.6 & 2732.7 \\
n=25 &  - & 2838.0 \\
 & & \\
$\Delta\nu$ & 105.5 $\pm$ 0.6 & 106.1 $\pm$ 0.4 \\ \hline 
\end{tabular}
\end{center}
\end{table}

\begin{figure*}
\resizebox{\hsize}{!}{\includegraphics{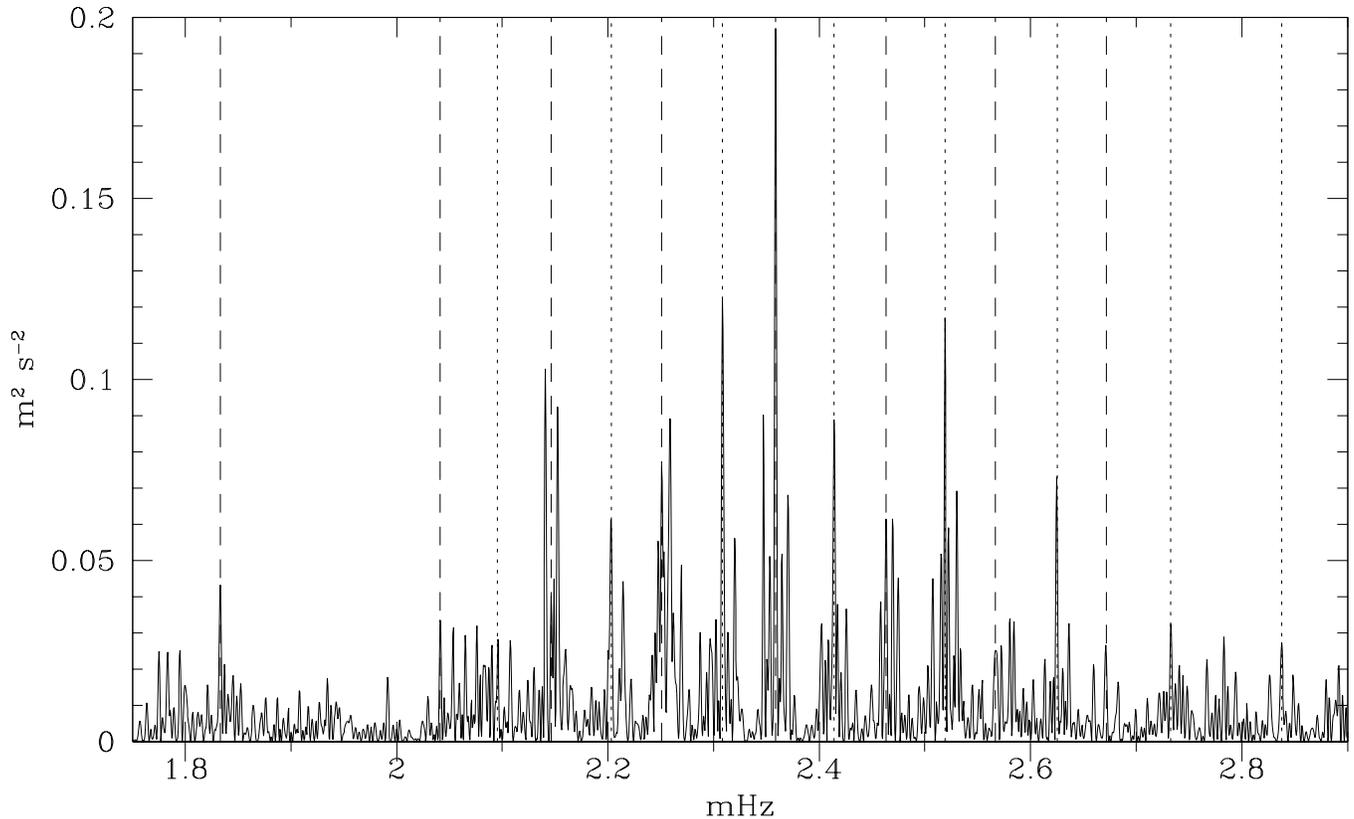}}
\caption{Identified p-mode oscillations of $\alpha$~Cen~A in the power 
spectrum. Dashed lines and dotted lines correspond respectively to the modes 
$l\,=\,0$ and $l\,=\,1$ listed in Table~2.}
\end{figure*}
   
\section{Concluding remarks and prospects}

Our observations of $\alpha$~Cen~A conduct to an obvious detection of p-mode 
oscillations. Several identifiable modes appear in the power spectrum between 
1.7 and 3 mHz with an average large splitting of 105.7 $\mu$Hz and an amplitude 
of about 35 \cms. These characteristics are in full agreement with the expected 
values scaling from the Sun presented in section 1 and will bring constraints 
on theoretical models.     

This result demonstrates the power of the simultaneous thorium RV method which 
reaches here the precision of 1 {\ms} at frequency higher than 0.5 mHz. 
We hope to obtain similar convincing seismological results on some bright 
solar-like stars with CORALIE. 

Observations of $\alpha$~Cen~A were conducted by T. Bedding and his collaborators 
during the same period with the UCLES spectrograph on the 3.9-m Anglo Australian
Telescope and the UVES spectrograph on the 8.2-m Very Large Telescope. We hope 
that these observations will conduct to multi-site data combination in order to
improve the spectral window and explore further the P-mode spectrum of
$\alpha$~Cen~A.

We are aware that ground based observation is 
far from the expected accuracy of the future space missions like MOST, MONS 
and COROT. However the future spectrograph HARPS (Pepe et al. \cite{pepe00}), 
which will be installed on 
the 3.6-m ESO telescope at La Silla Observatory at the end of 2002, is expected 
to observe stars with magnitude 5 dimmer than CORALIE (Bouchy et al. 
\cite{bouchy01}). It will thus be able to conduct complementary ground based 
asteroseismological study on a large sample of stars.

\begin{acknowledgements}
We are grateful to M. Mayor who encourages our program and gives us time
allocation at the Euler Swiss telescope. D. Queloz and L. Weber are acknowledged
for their help in the adaptation of the pipeline reduction needed for our 
observation sequences. This work was financially supported by the Swiss
National Science Foundation.    
\end{acknowledgements}

\end{document}